\documentclass[aps,prb,longbibliography,showpacs,twocolumn,superscriptaddress]{revtex4-2}

\usepackage[utf8]{inputenc}
\usepackage{amsmath,amsfonts,amssymb}
\usepackage{pifont}
\usepackage{rotating}
\usepackage{graphicx}
\usepackage{epstopdf}
\usepackage{color}
\usepackage[caption=false]{subfig}
\usepackage{soul}
\usepackage{booktabs}

\definecolor{dark-red}{rgb}{0.9,0.15,0.15}
\definecolor{dark-blue}{rgb}{0.15,0.15,0.4}
\definecolor{medium-blue}{rgb}{0,0,0.5}

\usepackage{hyperref}
\hypersetup{
	colorlinks=true,
	citecolor=blue,
	linkcolor=blue,
	urlcolor=blue
}

\begin{document}
\title{Low-temperature magnetism and spin dynamics in the disordered triangular-lattice Yb$^{3+}$ compound LiCaYb$_5$(BO$_3$)$_6$ }
\author{Monika Jawale}
\email{monikajawale12@gmail.com}
\affiliation{Department of Physics, Indian Institute of Technology Bombay, Mumbai 400076, India}
\author{Saikat Nandi}
\affiliation{Department of Physics, Indian Institute of Technology Bombay, Mumbai 400076, India}
\author{Prashanta K. Mukharjee}
\affiliation{Experimental Physics VI, Center for Electronic Correlations and Magnetism, University of Augsburg, D-86135 Augsburg, Germany}
\author{Philipp Gegenwart}
\affiliation{Experimental Physics VI, Center for Electronic Correlations and Magnetism, University of Augsburg, D-86159 Augsburg, Germany}
\author{A.V. Mahajan} 
\email{mahajan@phy.iitb.ac.in}
\affiliation{Department of Physics, Indian Institute of Technology Bombay, Mumbai 400076, India}

%-------------------------------
 %               ABSTRACT
 %%-------------------------------
\begin{abstract}
	The interplay between geometric frustration, spin--orbit coupling, and structural disorder can give rise to unconventional magnetic ground states in rare-earth triangular-lattice magnets. We report low-temperature magnetic and $^7$Li NMR investigations of the disordered triangular-lattice compound LiCaYb$_5$(BO$_3$)$_6$ (LCYBO). Rietveld refinement confirms a hexagonal $P6_522$ structure with partial Ca/Yb antisite disorder and fractional Li occupancy. Magnetic susceptibility and magnetization measurements indicate a well-isolated effective $J_{\mathrm{eff}}=1/2$ Kramers doublet with weak antiferromagnetic interactions ($\theta_{\mathrm{CW}} \approx -0.54$~K). The specific heat reveals a weak anomaly near 0.43 K, suggesting the development of short-range correlated magnetism rather than conventional long-range order. $^7$Li NMR spectra broaden strongly upon cooling, consistent with increasingly inhomogeneous internal magnetic fields. The spin-lattice relaxation exhibits two relaxation components consistent with disorder-induced distributions of local magnetic environments arising from antisite disorder and competing exchange pathways. Our results establish LCYBO as a structurally disordered frustrated triangular-lattice magnet with correlated low-energy spin dynamics.
\end{abstract}
				
\date{\today}

\maketitle

\section{Introduction}
Magnetic frustration occurs when the geometry of a lattice or competing exchange interactions prevent all pairwise spin interactions from being simultaneously satisfied. As a consequence, the system cannot easily select a unique classical ground state and instead develops a manifold of nearly degenerate configurations. Such degeneracy enhances the role of both thermal and quantum fluctuations, often leading to magnetic behavior that differs markedly from that of conventional ordered magnets. Understanding how frustration affects the stability of long-range magnetic order, the emergence of short-range correlations, and the formation of exotic ground states remains a central topic in modern condensed-matter physics~\cite{diep2013, balents2010, savary2016}.

Geometrically frustrated magnets based on triangular lattices with antiferromagnetic (AFM) interactions provide an important platform for exploring these phenomena~\cite{anderson1973}. In an ideal triangular arrangement, spins cannot simultaneously satisfy all AFM bonds, resulting in strong geometric frustration. For the nearest-neighbor Heisenberg antiferromagnet on a triangular lattice, the classical ground state adopts a noncollinear $120^{\circ}$ spin structure. Although quantum fluctuations renormalize the ordered moment, long-range $120^{\circ}$ order remains stable for the $S=1/2$ nearest-neighbor Heisenberg model~\cite{Huse1988,Bernu1994}. In the presence of additional effects such as exchange anisotropy, competing interactions, or disorder, frustration can suppress conventional long-range magnetic order and stabilize unconventional states, including quantum spin liquids (QSLs), spin glasses, and partially ordered phases~\cite{li2020}. 

Rare-earth magnets form a particularly rich platform for exploring these phenomena because strong spin–orbit coupling (SOC) and crystal electric field (CEF) effects reduce the complex $4f$ electronic structure to an effective low-energy Kramers doublet~\cite{rau2019, li2020}. In the low-temperature limit, these degrees of freedom can be modeled as effective spin-$\tfrac{1}{2}$ moments with anisotropic $g$-factors and direction-dependent exchange interactions arising from spin–orbit entanglement of spin and orbital degrees of freedom. Among rare-earth ions, Yb$^{3+}$ is especially appealing because its ground-state Kramers doublet is protected by time-reversal symmetry and is typically well separated from excited CEF states. Consequently, Yb-based magnets often display clear signatures of anisotropic exchange interactions and provide ideal systems to study the interplay between frustration, dimensionality, and quantum fluctuations. For Yb$^{3+}$ ($4f^{13}$; $L=3$, $S=\tfrac{1}{2}$, $J=\tfrac{7}{2}$), the eightfold-degenerate $^{2}F_{7/2}$ manifold is split by non-cubic CEFs into four Kramers doublets, with the lowest one serving as a robust pseudospin-$\tfrac{1}{2}$ state. The exchange interactions between Yb moments are typically weak, of the order of a few kelvins, and therefore low-temperature measurements are essential for probing their intrinsic magnetic behavior.

Over the past decade, Yb-based triangular magnets have played an   important role in the study of frustrated magnetism. Compounds such as YbMgGaO$_4$ have been proposed as QSL candidates due to the absence of magnetic ordering down to millikelvin temperatures, although significant debate persists regarding the impact of structural disorder, particularly Mg/Ga site mixing~\cite{li2015,li2017, li2020}. A related system, YbZnGaO$_4$, shows disorder-driven spin-glass-like freezing attributed to Zn/Ga mixing~\cite{zhang2018,ma2020}. Other triangular-lattice Yb systems, including NaYbO$_2$~\cite{ding2019}, NaBaYb(BO$_3$)$_2$~\cite{guo2019}, KBaYb(BO$_3$)$_2$ and YbBO$_3$~\cite{Sala2023}, show very low ordering temperatures. More recently, YbZn$_2$GaO$_5$ has been proposed as a triangular-lattice system with possible Dirac-type QSL behavior~\cite{Bag2024}. Collectively, these systems demonstrate how strong SOC, geometric frustration, and structural disorder can bring about a variety of correlated magnetic ground states. Layered borates containing Yb$^{3+}$ ions have recently emerged as another promising class of frustrated magnets. Their structures contain networks of BO$_3$ triangles that mediate superexchange pathways between Yb ions, frequently leading to triangular or distorted triangular magnetic lattices. These borates often contain multiple crystallographic Yb sites and may also exhibit cation disorder involving alkali or alkaline-earth ions. Such structural characteristics provide avenues for bond randomness and exchange anisotropy, making borate systems excellent platforms for studying how local distortions and disorder affect frustrated magnetism~\cite{somesh2023, bag2021, guchhait2025, zeng2020, bag2021, khatua2022}.

In this work, we investigate the compound LiCaYb$_5$(BO$_3$)$_6$ (LCYBO), a Yb-based triangular-lattice system whose structural details were reported in Ref.~\cite{Gao2021}. LCYBO crystallizes in the hexagonal space group $P6_522$ and contains three distinct Yb sites that together form a slightly distorted triangular network. Structural analysis shows that two of these Yb sites exhibit site mixing with Ca, whereas the Li crystallographic site is only half-occupied, indicating the presence of intrinsic vacancies. Such antisite disorder and incomplete Li occupancy can locally modify the Yb–Yb exchange pathways and introduce randomness into the magnetic coupling strengths. Therefore, LCYBO provides an opportunity to examine the combined effects of geometric frustration, exchange anisotropy, and structural disorder on low-energy magnetic correlations. %Therefore, LCYBO provides an opportunity to examine how intrinsic geometric frustration, combined with disorder-induced exchange randomness, shapes the magnetic ground state. 
To address this, we employ a combination of bulk and local experimental probes. Magnetization measurements are used to determine the effective magnetic moment, extract the sign and strength of the dominant exchange interactions, and verify the presence of an isolated Kramers doublet. Heat-capacity measurements reveal thermodynamic anomalies and quantify the magnetic entropy released near the transition. In addition, $^{7}$Li NMR acts as a sensitive local probe of internal magnetic fields, enabling us to monitor the buildup of short-range correlations, determine the hyperfine coupling, and examine the slowing down of spin dynamics.  Given the presence of Yb$^{3+}$ Kramers doublets and the strong SOC/CEF-driven anisotropy, LiCaYb$_5$(BO$_3$)$_6$ may exhibit rich field- and temperature-dependent magnetic behavior arising from the interplay of frustration, quantum fluctuations, and anti-site disorder. Experimental studies of this system can therefore provide insight into anisotropic triangular-lattice magnetism and the evolution from weakly ordered states toward quantum-disordered regimes in frustrated rare-earth magnets.
 
\section {Experimental Details}

Polycrystalline LiCaYb$_5$(BO$_3$)$_6$ (LCYBO) was synthesized by a conventional solid-state reaction method~\cite{Gao2021}. High-purity starting materials were used: Li$_2$CO$_3$ (Alfa Aesar, 99.9\%; 5\% excess to compensate for volatilization), CaCO$_3$ (Aldrich, 99\%), Yb$_2$O$_3$ (Alfa Aesar, 99.995\%), and H$_3$BO$_3$ (Alfa Aesar, 99.9\%; 10\% excess to compensate for volatilization). The powders were thoroughly mixed in an agate mortar, pelletized, and heated in air at 600~$^\circ$C for 10~h in a box furnace. After furnace cooling, the pellet was crushed, reground, and repelletized, followed by a second heat treatment at 850~$^\circ$C for an additional 10~h. The final product was ground into a fine powder for structural and physical characterization.

Phase purity was confirmed by powder x-ray diffraction measurements performed at room temperature using a Rigaku diffractometer (Smartlab 9~kW) with Cu K$_\alpha$ radiation ($\lambda = 1.5406$~\AA). Magnetic susceptibility was measured as a function of temperature ($0.4 \leq T \leq 300$~K) in applied fields of $H = 100$~Oe and 10~kOe. Isothermal magnetization [$M(H)$] measurements were also performed at various temperatures. Low-temperature magnetization measurements down to 0.4~K were carried out using a SQUID vibrating sample magnetometer equipped with a $^3$He insert (iQuantumHe3, Quantum Design). Specific-heat measurements were performed using the standard relaxation technique in a Physical Property Measurement System (PPMS, Quantum Design) under magnetic fields $0 \leq H \leq 90$~kOe. The contribution from the addenda was measured separately and subtracted from the raw data. Additional low-temperature specific-heat measurements down to 0.4~K were carried out at magnetic fields of 0 and 20~kOe using a $^3$He option in a PPMS Dynacool (Quantum Design). $^{7}$Li nuclear magnetic resonance (NMR) measurements were performed using a Redstone Tecmag spectrometer in a swept-field magnet (Cryomagnetics Inc.) at a fixed frequency of 100~MHz down to 4~K. Field-sweep NMR spectra were recorded at various temperatures. 

\section {Results and discussion}
\subsection{Structural analysis}
\begin{figure}[ht]
	\centering
	\includegraphics[width=1\linewidth]{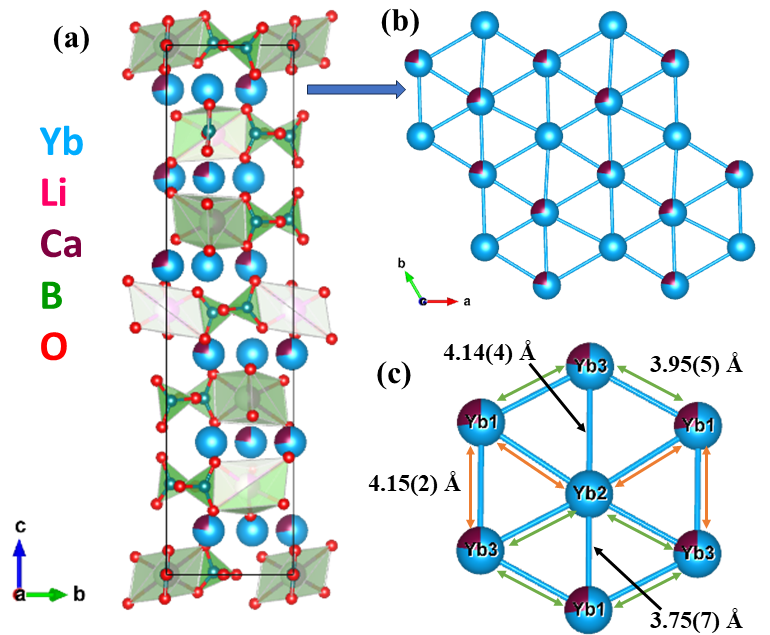}
	\caption{(a) Crystal structure of LCYBO showing the stacking of the Yb-containing layers along the $c$-axis. Yb(1), Yb(2), and Yb(3) sites, which belong to the same distorted triangular magnetic layer, are shown together with Li, Ca, B, and O in blue, pink, maroon, green, and red, respectively; the Ca sites are partially obscured in this projection. Equivalent distorted triangular magnetic layers are stacked along the $c$-axis. (b) $ab$-plane projection of a single distorted triangular magnetic layer formed by the Yb1, Yb2, and Yb3 sites. (c) Distorted triangular arrangement of the Yb ions within the same magnetic layer, showing the inequivalent Yb--Yb distances (indicated). Structure visualization was performed using VESTA~\cite{vesta}.}
	\label{yb_CS}
\end{figure}

\begin{figure}[ht]
	\centering
	\includegraphics[width=0.9\linewidth]{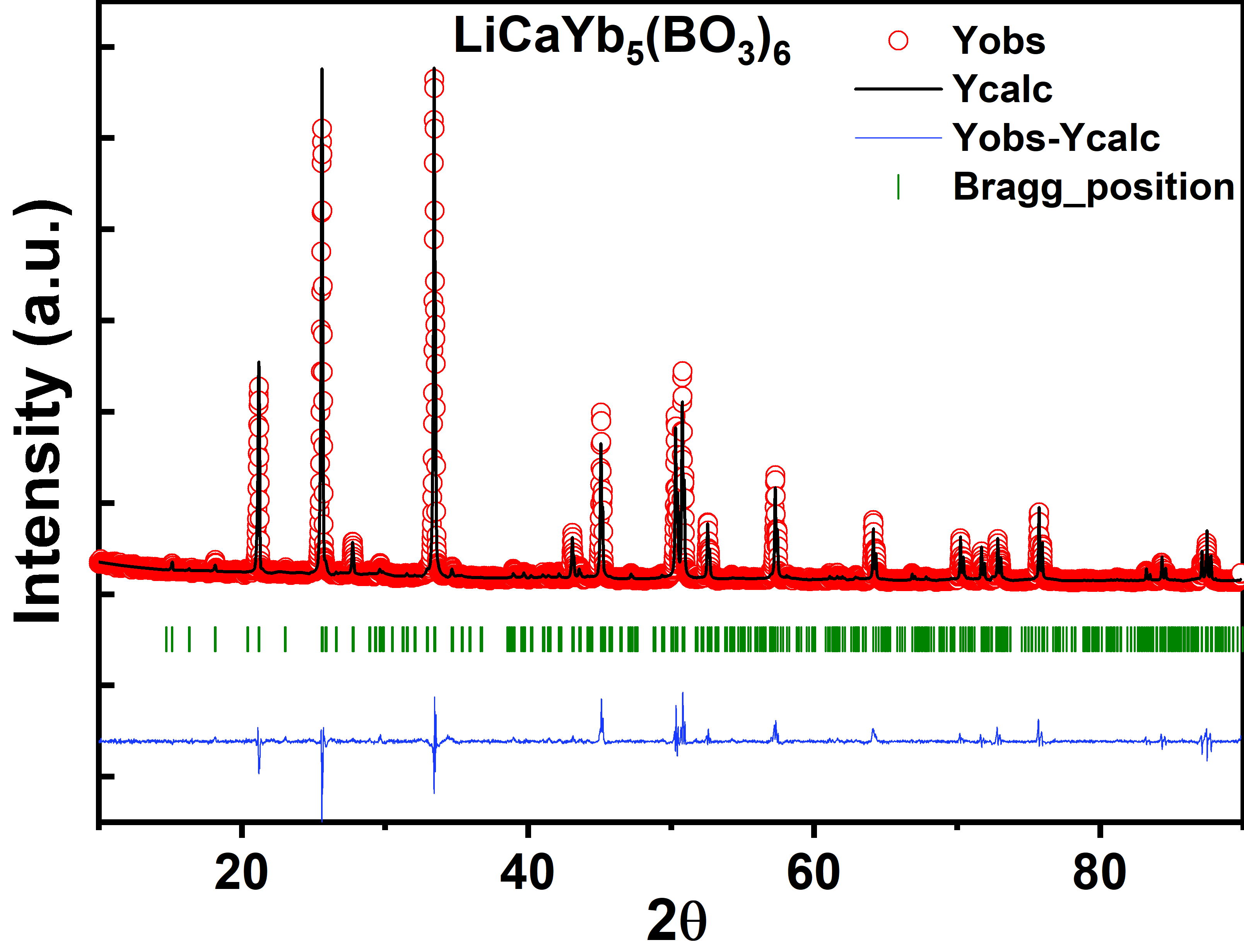}
	\caption{Rietveld refinement of the powder XRD pattern of LCYBO showing observed, calculated, and difference profiles along with Bragg peak positions.}
	\label{yb_xrd}
\end{figure}

Powder x-ray diffraction (XRD) patterns were collected at room temperature. Rietveld refinement confirms that LCYBO crystallizes as a single phase within the detection limit of the measurement, as shown in Fig.~\ref{yb_xrd}. The refinement yields lattice parameters $a = b = 6.9588(2)$~\AA, $c = 25.1676(4)$~\AA, $\alpha = \beta = 90^{\circ}$, and $\gamma = 120^{\circ}$, consistent with the hexagonal space group $P6_522$. The refined crystal structure is illustrated in Fig.~\ref{yb_CS}, and the atomic positions are listed in Table~\ref{tab: table5}. The refinement quality is characterized by $\chi^2 = 3.9$, $R_p = 25.6$, $R_{wp} = 20.6$, and $R_{exp} = 10.3$. The obtained lattice parameters are in good agreement with previously reported values~\cite{Gao2021}.

LCYBO contains three crystallographically distinct Yb sites, denoted Yb(1), Yb(2), and Yb(3). Yb(1) and Yb(3) are coordinated by six oxygen atoms, forming YbO$_6$ octahedra, whereas Yb(2) is eightfold coordinated. The Rietveld refinement indicates Ca/Yb antisite mixing at the Yb(1) and Yb(3) sites, with Ca occupying approximately 25\% of each site, while the Yb(2) site remains fully occupied by Yb$^{3+}$. The Li site is tetrahedrally coordinated by oxygen but exhibits partial occupancy (occupancy $\approx 0.5$), indicating intrinsic Li vacancies. Boron atoms (B1 and B2) are threefold coordinated and form planar BO$_3$ triangles. The Yb(1)O$_6$, Yb(2)O$_8$, and Yb(3)O$_6$ polyhedra are linked through shared edges and corners, forming extended metal–oxygen layers in the $ab$ plane. These layers are separated along the $c$ axis by LiO$_4$ tetrahedra and BO$_3$ units. Within each layer, the Yb ions form a distorted triangular lattice. Yb(1), Yb(2), and Yb(3) are crystallographically inequivalent sites within the same magnetic network and should not be considered as independent magnetic subsystems. The layers shown in Fig.~\ref{yb_CS}(a) are related through the crystal symmetry and stacking. Thus, LCYBO forms a connected Yb magnetic network, although the three Yb sites may have different local environments, exchange couplings, and $g$ tensors.
Within each layer the Yb–Yb distances are 3.75(7)~\AA, 3.95(5)~\AA, 4.14(4)~\AA, and 4.15(2)~\AA\ [Fig.~\ref{yb_CS}(c)]. The variation in Yb--Yb distances and the associated Yb--O--B--O--Yb pathways may lead to nonuniform exchange interactions. However, exchange is not determined by Yb--Yb distance alone and also depends on bond angles, orbital overlap, and the local CEF environment; therefore, the shortest 3.75(7)~\AA\ bond does not necessarily correspond to the strongest exchange. The average intralayer separation is $d_{\mathrm{intra}} \approx 4.016$~\AA. Successive Yb layers are stacked along the $c$ axis with an interlayer spacing of $d_{\mathrm{inter}} \approx 4.19$~\AA. The ratio $d_{\mathrm{inter}}/d_{\mathrm{intra}} \approx 1.04$ indicates that the interlayer separation is only slightly larger than the intralayer distance. This suggests a weakly distorted layered triangular magnetic lattice in which both in-plane and interlayer exchange interactions may play an important role. For comparison, the ratios of interlayer to intralayer distances for several Yb-based magnets, along with their corresponding magnetic ground states, are summarized in Table~\ref{tab:yb_comparison}. The combination of distorted triangular geometry, partial Ca/Yb antisite mixing at the Yb1 and Yb3 sites, and intrinsic Li-site vacancies introduces local variations in Yb–O coordination and modifies the Yb–Yb exchange pathways. Such structural disorder can generate bond randomness and spatially varying exchange interactions within the triangular lattice. The present structural data, however, do not allow the relative contributions of bond randomness, exchange anisotropy, and geometric frustration to be quantified independently. Overall, LCYBO hosts a geometrically frustrated Yb$^{3+}$ magnetic sublattice embedded in a nonmagnetic borate framework. The coexistence of triangular geometry and structural disorder makes this system an interesting platform for investigating the interplay between frustration, exchange anisotropy, and disorder in rare-earth magnets.

%%%%%%%%%%%%%%%%%%%%%%%%%%%%%%%%%%%%%%%%%%%%%%%%
%%%%%%%%%%%%%%%%%%%%%%%%%%%%%%%%%%%%%%%%%%%%%%%% 

%%%%%%%%%%%%%%%%%%%%%%%%%%%%%%%%%%%%%%%%%%%%%%%%
%%%%%%%%%%%%%%%%%%%%%%%%%%%%%%%%%%%%%%%%%%%%%%%%

%%%%%%%%%%%%%%%%%%%%%%%%%%%%%%%%%%%%%%%%%%%%%%%%
\begin{table}[h]
	\caption{Refined atomic positions of LiCaYb$_{5}$(BO$_{3}$)$_{6}$.}
	\label{tab: table5}
	\centering
	\begin{tabular}{c c c c c c c}
		\toprule
		Atom & Wyckoff& x & y & z & $B_{iso}$ & Occupancy \tabularnewline
		  & position & (\AA) & (\AA) &  (\AA) &  (\AA$^{2}$) &  \tabularnewline
		\midrule
		\hline
		Yb(2) & 6b & 0.3256 & 0.6743 & 0.0833 & 1.6818 & 1.000 \tabularnewline
		Ca(1) & 6b & 0.01068 & -0.01068 & 0.0833 & 1.7213 & 0.263 \tabularnewline
		Yb(1) & 6b & 0.01068 & -0.01068 & 0.0833 & 1.7213 & 0.737 \tabularnewline
		Ca(2) & 6b & 0.66593 & 0.33407 & 0.0833 & 1.7765 & 0.763 \tabularnewline
		Yb(3) & 6b & 0.66593 & 0.33407 & 0.0833 & 1.7765 & 0.237 \tabularnewline
		B(1)  & 12c & 0.30106 & 0.2574 & 0.00869 & 2.138  & 1.000 \tabularnewline
		B(2)  & 6a  & 0.56966 & 0.0000 & 0.0000  & 2.138  & 1.000 \tabularnewline
		Li  & 6a  & 0.2349  & 0.0000 & 0.0000  & 1.5791 & 0.500 \tabularnewline
		O(1)  & 12c & 0.4534  & 0.6628 & 0.1653  & 1.0264 & 1.000 \tabularnewline
		O(2)  & 6a  & 0.8684  & 0.0000 & 0.0000  & 1.0264 & 1.000 \tabularnewline
		O(3)  & 12c & 0.1087  & 0.7771 & 0.0295  & 1.0264 & 1.000 \tabularnewline
		O(4)  & 12c & 0.6910  & 0.0262 & 0.0484  & 1.0264 & 1.000 \tabularnewline
		O(5)  & 12c & 0.3558  & 0.2867 & 0.0483  & 1.0264 & 1.000 \tabularnewline
		\bottomrule  
	\end{tabular}
\end{table}
%%%%%%%%%%%%%%%%%%%%%%%%%%%%%%%%%%%%%%%%%%%%%%%%
\subsection{DC magnetization}

\begin{figure}[ht]
	\centering
	\includegraphics[width=0.9\linewidth]{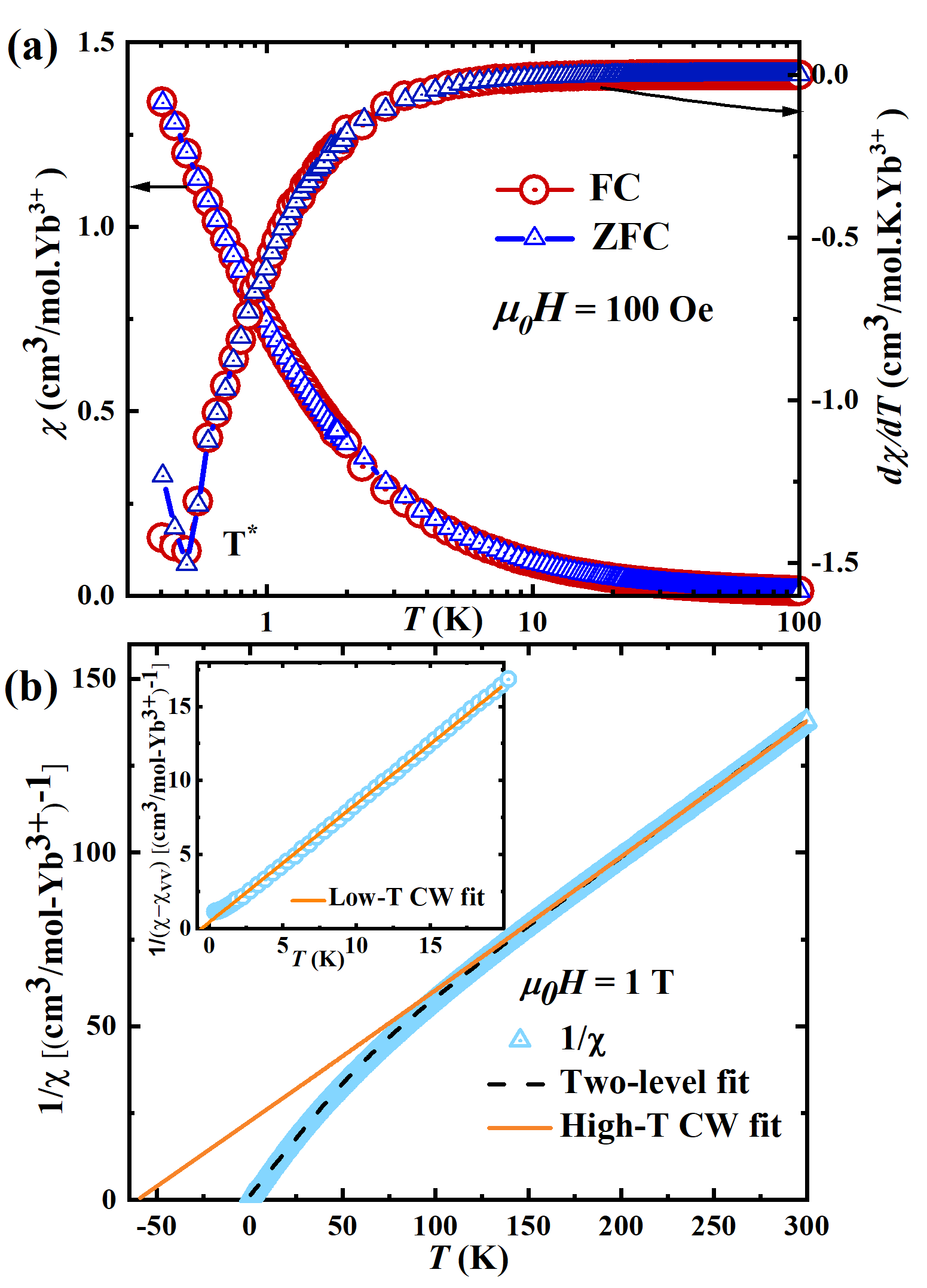}
	\caption{(a)Temperature-dependent magnetic susceptibility of LCYBO measured in an applied field of 100~Oe under ZFC and FC conditions (left axis), and its first derivative (right axis) showing a clear minimum near 0.5~K. (b) Inverse magnetic susceptibility measured at 10~kOe under ZFC conditions for LCYBO. The dashed and solid lines represent the two-level fit and the high-temperature Curie–Weiss fit (180–300~K), respectively. Inset: Curie–Weiss fit in the low-temperature range (0.75–20~K).}
	\label{yb_mt}
\end{figure}

To investigate the magnetic behavior of LCYBO, we performed temperature-dependent DC magnetic susceptibility measurements in the range $0.4~\text{K} \leq T \leq 300~\text{K}$ and isothermal magnetization measurements up to $\mu_{0}H = 70~\text{kOe}$. The low-field susceptibility measured at 100~Oe is shown in Fig.~\ref{yb_mt}(a). Although no visible anomaly appears in the raw $\chi(T)$ curve (left axis of Fig.~\ref{yb_mt}(a)), the temperature derivative $d\chi/dT$ (right axis of Fig.~\ref{yb_mt}(a)) reveals a clear minimum near $\sim 0.5$~K, indicating the onset of magnetic correlations. Furthermore, the zero-field-cooled (ZFC) and field-cooled (FC) $\chi(T)$ curves do not bifurcate down to 0.4~K, arguing against canonical spin-glass behavior. A weak separation is observed only in $d\chi/dT$, but its small magnitude suggests it does not indicate a true spin-freezing transition.

At higher temperatures ($T > 180$~K), the inverse magnetic susceptibility, $1/\chi(T)$ data measured in ZFC mode at 10~kOe (Fig.~\ref{yb_mt}(b)) is described by a modified Curie–Weiss (CW) law:
\begin{equation}
	\chi(T) = \chi_{0} + \frac{C}{T - \theta_{\mathrm{CW}}},
	\label{m}
\end{equation}
where $\chi_{0}$ is the temperature-independent contribution (core diamagnetism $+$ Van Vleck paramagnetism). $C$ is the Curie constant and $\theta_{\mathrm{CW}}$ is the Curie-Weiss temperature. The Curie constant is related to the effective moment by $C = N_A\mu_{\mathrm{eff}}^2/3k_B$, and $\mu_{\mathrm{eff}}=g\sqrt{J(J+1)}\mu_B$, where $N_A$ is Avogadro’s number, $g$ is the Landé $g$-factor, $\mu_B$ is the Bohr magneton, and $J$ is the total angular momentum.

The high-temperature fit in the range 180 - 300~K yields $\chi_{0}^{HT} = -2.2(2) \times 10^{-4}$~cm$^3$/mol, $C^{HT} = 2.69(1)$~cm$^3$K/mol and $\theta_{\mathrm{CW}}^{HT} = -60.4(1)$~K. The extracted effective moment $\mu_{\mathrm{eff}}^{HT} = 4.64~\mu_B$/Yb is close to the free-ion value $4.54~\mu_B$ expected for Yb$^{3+}$. The large and negative \(\theta_{\mathrm{CW}}^{\mathrm{HT}}\) primarily reflects contributions from excited CEF levels and Van Vleck susceptibility.

A change in the slope of $1/\chi(T)$ below $\sim$ 100~K signals depopulation of the excited CEF levels, as commonly seen in Yb-based materials~\cite{Ranjith2019, somesh2023, Sebastian2025, arjun2023, singh2024, sarkar2019, ding2019}. To isolate the contribution from the lowest-energy state, we subtract the temperature-independent Van Vleck term $\chi_{\mathrm{VV}}$, and replotted $1/(\chi-\chi_{\mathrm{VV}})$ (inset of Fig.~\ref{yb_mt}(b)). The value of $\chi_{\mathrm{VV}}$ was obtained from the high-field linear slope of the 0.4~K magnetization (see below).

A low-temperature Curie–Weiss fit in the range 0.75–20~K gives $C^{LT} = 1.247(3)$~cm$^3$K/mol, $\theta_{\mathrm{CW}}^{LT} = -0.54(2)$~K, from which we obtain $\mu_{\mathrm{eff}}^{LT} = 3.159(4)~\mu_B$. This value is consistent with an effective $J_{\mathrm{eff}} = 1/2$ Kramers doublet (expected $\mu_{\mathrm{eff}} = 3.2~\mu_B$). The corresponding average $g$-factor is $g_{\mathrm{avg}} = 3.647(5)$, which is slightly larger than the free-spin value, indicating the presence of spin-orbit entanglement and anisotropy. Since the bulk susceptibility probes all crystallographically distinct Yb sites simultaneously, the extracted $\mu_{\mathrm{eff}}^{LT}$ and $g_{\mathrm{avg}}$ represent average magnetic parameters of the Yb sublattice and do not imply identical local moments or $g$ tensors at the three Yb sites. The small magnitude of $\theta_{\mathrm{CW}}^{LT}$, suggests weak, predominantly antiferromagnetic exchange interactions between Yb$^{3+}$ moments.

The effect of the CEF on the susceptibility was further analysed by fitting the data above 25~K with a simple two-level model~\cite{Mugiraneza2022,arjun2023}:
\begin{equation}
	\chi(T) = \chi_0 + \frac{1}{8(T - \theta_{\mathrm{CW}})} 
	\left(\frac{\mu_{\mathrm{eff,1}}^2 + \mu_{\mathrm{eff,2}}^2 e^{-\Delta_{\mathrm{CEF}}/T}}
	{1 + e^{-\Delta_{\mathrm{CEF}}/T}} \right),
	\label{m1}
\end{equation}
where $\Delta_{\mathrm{CEF}}$ is the energy gap between the ground and first excited CEF states, and $\mu_{\mathrm{eff,1}}$, $\mu_{\mathrm{eff,2}}$ are the corresponding effective moments. Fitting the data above 25~K (Fig.~\ref{yb_mt}(b)) yields $\Delta_{\mathrm{CEF}} = 200.4$~K, $\theta_{\mathrm{CW}} = -1.8$~K, $\mu_{\mathrm{eff,1}} = 3.47~\mu_B$, and $\mu_{\mathrm{eff,2}} = 5.29~\mu_B$. The large $\Delta_{\mathrm{CEF}}$ indicates a well-isolated ground-state doublet, consistent with other Yb-based magnets~\cite{arjun2023, singh2024, sarkar2019, ding2019}. Since this two-level approach is qualitative, determination of the full CEF scheme requires spectroscopic probes such as inelastic neutron scattering or Raman scattering.

\begin{figure}[ht]
	\centering
	\includegraphics[width=0.9\linewidth]{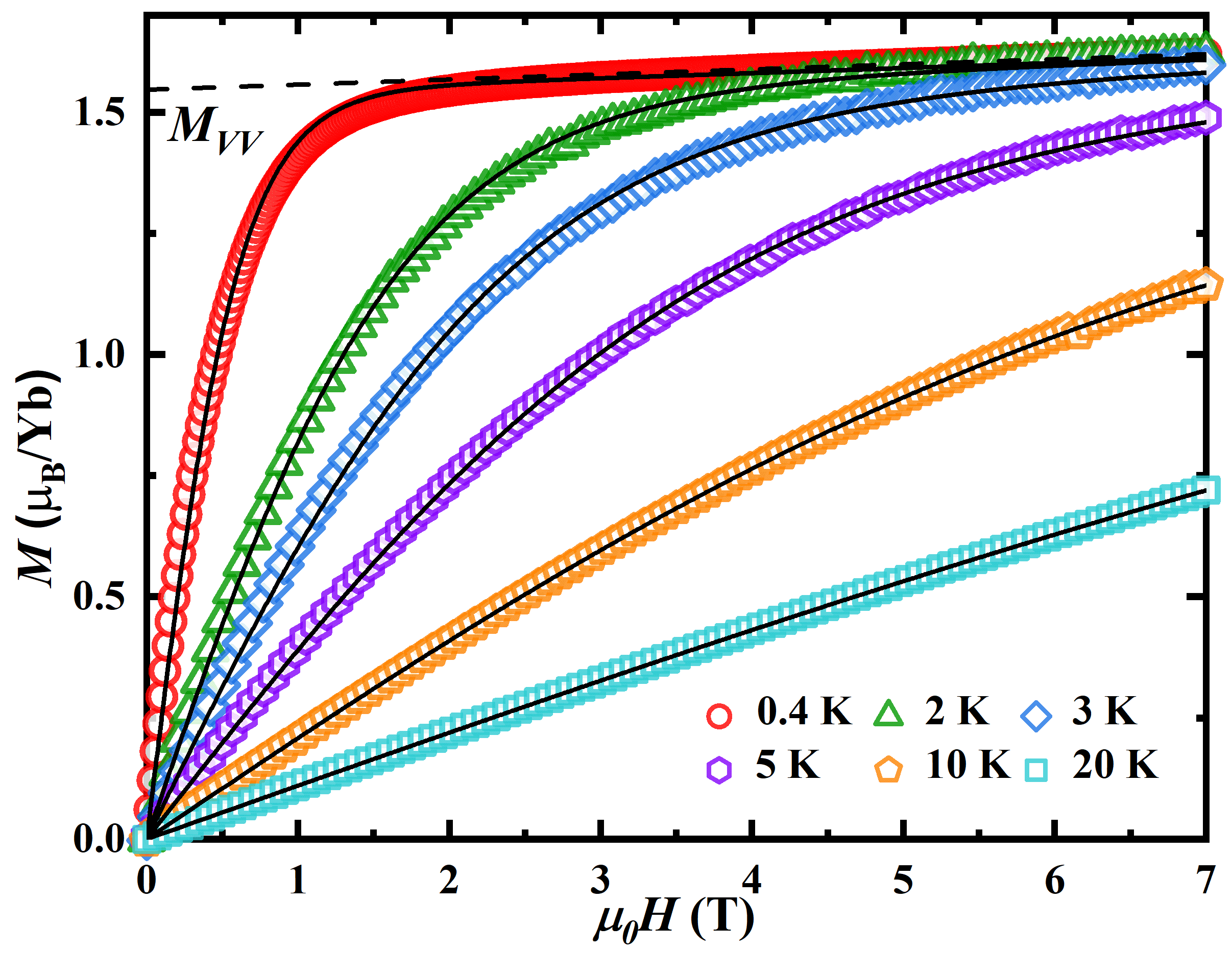}
	\caption{Isothermal magnetization $M(H)$ measured at 0.4 K, 2K and 50 K for LiCaYb$_{5}$(BO$_{3}$)$_{6}$. Dashed line: linear fit extrapolated to saturation magnetization; solid line: Brillouin function fit. }
	\label{yb_m3}
\end{figure}

Isothermal magnetization $M(H)$ measured at several temperatures is shown in Fig.~\ref{yb_m3}. At 0.4~K, the high-field part of $M(H)$ becomes linear (for $\mu_{0}H > 5$~T), from which we extract the temperature-independent Van Vleck susceptibility $\chi_{\mathrm{VV}} = 5.86(2) \times 10^{-3}$~cm$^3$/mol. This relatively large $\chi_{\mathrm{VV}}$ is typical for Yb$^{3+}$ compounds and reflects strong SOC and moderate CEF gaps that produce substantial virtual admixing of excited CEF states. The intercept of the linear high-field fit gives a saturation moment $M_s = 1.55~\mu_B$/Yb, corresponding to $g_{\mathrm{avg}} \approx 3.09$, which is consistent with the low-temperature Curie analysis.

To model the low-temperature magnetization, we used an expression that combines the Van Vleck background with a Brillouin contribution for weakly interacting $J_{\mathrm{eff}} = 1/2$ moments:
\begin{equation}
	M(H) = \chi_{\mathrm{VV}} H + gJ_{\mathrm{eff}}N_A\mu_B B_{J_{\mathrm{eff}}}(H, T),
	\label{eq:MH}
\end{equation}
where $B_{J_{\mathrm{eff}}}(H, T)$ is the Brillouin function for $J_{\mathrm{eff}} = 1/2$~\cite{kittel2018}. In the fits we fixed $J_{\mathrm{eff}} = 1/2$ and $\chi_{\mathrm{VV}}$ while allowing $g$ to refine. The Brillouin+Van-Vleck model reproduces the measured isotherms well at low temperature, consistent with a picture of weakly interacting effective spin-$\tfrac{1}{2}$ moments.

\subsection{Specific Heat}

\begin{figure*}
	\centering
	\includegraphics[width=1\linewidth]{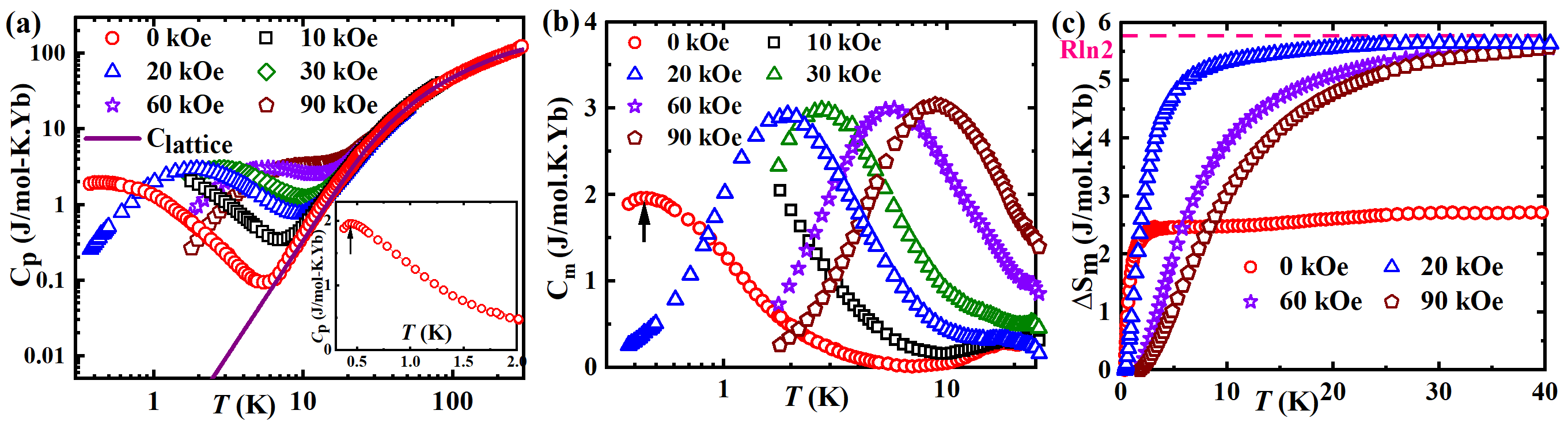}
	\caption{(a)Temperature and field dependence of the total specific heat of LCYBO plotted on a log–log scale to emphasize the low-$T$ region.  The solid line represents the estimated lattice contribution. The inset shows the low-temperature zero-field data below 2~K, where the arrow marks the anomaly near 0.43~K.  
	(b)  Magnetic contribution to the specific heat, $C{\text{m}}(T)$, obtained after subtracting the lattice contribution. The broad maximum shifts toward higher temperature with increasing magnetic field, indicating field-induced evolution of the low-energy magnetic excitations. The arrow marks the low-temperature anomaly in zero field.
	(c) Magnetic entropy change, $\Delta S{\text{m}}(T)$, obtained by integrating $C{\text{m}}/T$. The dashed line indicates the expected entropy value $R\ln2$ for an effective spin-$1/2$ ground-state doublet of Yb$^{3+}$.}
	\label{yb_cp}
\end{figure*}

To investigate the low-energy excitations and magnetic ground state of LCYBO, we measured the specific heat at constant pressure, $C_{p}(T)$, in the temperature range 0.4--300~K in zero field and 2--50~K under applied magnetic fields up to 9~T (for 2~T, measurements were performed down to 0.4~K). The total specific heat is shown in Fig.~\ref{yb_cp}(a). In zero field, $C_{p}(T)$ exhibits a weak upturn below $\sim 5$~K, signaling the development of short-range magnetic correlations. One can see a broad feature centered near 0.43~K [inset of Fig.~\ref{yb_cp}(a)], which is in contrast to a sharp $\lambda$-type signature expected for conventional long-range ordering. Such a broadened feature may arise from the combined effects of structural disorder, a distribution of exchange interactions, and intrinsic frustration associated with the distorted triangular geometry. Nevertheless, this behavior is consistent with the development of correlated low-temperature magnetic behavior among the Yb$^{3+}$ moments. Since the measurements extend only down to 0.4 K, the observed feature cannot by itself distinguish between a broadened magnetic transition, short-range or crossover behavior, a Schottky contribution, or disorder-induced internal fields.

Upon applying a magnetic field, the anomaly at 0.43~K is progressively suppressed and evolves into a broad maximum centered around 2~K for 20~kOe. Such behavior is consistent with a Schottky-type anomaly arising from Zeeman splitting of the ground-state Kramers doublet, as commonly observed in Yb-based magnets with an effective $J_{\rm eff}=1/2$ ground state. However, the zero-field anomaly may contain contributions from low-energy magnetic correlations and other effects such as disorder-induced internal fields.

At high temperatures, the specific heat is dominated by the lattice (phonon) contribution $C_{\mathrm{ph}}(T)$. To extract the magnetic contribution $C_{\text{m}}(T)$, the zero-field specific heat $C_{\text{p}}(T)$ in the temperature range 40--200\,K was modeled using a combination of one Debye and two Einstein modes (1D+2E), which approximate the contributions from acoustic and optical phonon branches of the lattice. The Debye and Einstein contributions are given by

\begin{equation}
	C_{\mathrm{Debye}}(T) = C_{\mathrm{D}} \, 9R \left(\frac{T}{\theta_{\mathrm{D}}}\right)^{3}
	\int_{0}^{x_{\mathrm{D}}} \frac{x^{4} e^{x}}{(e^{x}-1)^{2}} \, dx ,
	\label{eq:Debye1}
\end{equation}

\begin{equation}
	C_{\mathrm{Einstein}}(T) = \sum_{j=1}^{2} C_{\mathrm{E},j} \, 3R
	\left(\frac{\theta_{\mathrm{E},j}}{T}\right)^{2}
	\frac{\exp(\theta_{\mathrm{E},j}/T)}{\big(\exp(\theta_{\mathrm{E},j}/T)-1\big)^{2}},
	\label{eq:Einstein1}
\end{equation}

where $\theta_{\mathrm{D}}$ and $\theta_{\mathrm{E},j}$ ($j=1,2$) denote the Debye and Einstein temperatures, $C_{\mathrm{D}}$ and $C_{\mathrm{E},j}$ are the corresponding weighting factors, and $R$ is the universal gas constant. The weighting factors were constrained such that their sum is close to unity. Since the formula unit contains 31 atoms, the lattice possesses a total of $3N = 93$ vibrational modes consisting of three acoustic modes and 90 optical modes. The Debye term represents the acoustic phonon contribution, while the Einstein terms account for the optical phonon modes. The fitting yields the weighting factors 
$C_{\mathrm{D}} = 0.205 \pm 0.04$, 
$C_{\mathrm{E}_1} = 0.434 \pm 0.05$, and 
$C_{\mathrm{E}_2} = 0.442 \pm 0.01$. 
The corresponding characteristic temperatures are 
$\theta_{\mathrm{D}} = 194(21)$\,K, 
$\theta_{\mathrm{E}_1} = 326(29)$\,K, and 
$\theta_{\mathrm{E}_2} = 766(29)$\,K.
The resulting lattice contribution $C_{\mathrm{ph}}(T)$ was extrapolated to low temperatures using a $\beta T^{3}$ dependence. The magnetic specific heat $C_{\text{m}}(T)$ was then obtained by subtracting the lattice contribution from the total specific heat, as shown in Fig.~\ref{yb_cp}(a). The zero-field magnetic specific heat $C_{\text{m}}(T)$ exhibits an anomaly near $T^{*} \approx 0.43$\,K, which shifts to higher temperatures upon applying a magnetic field. This field-dependent peak is consistent with a Schottky-type anomaly arising from the Zeeman splitting of the ground-state Kramers doublet of Yb$^{3+}$, similar to behavior observed in other Yb-based magnetic systems~\cite{arjun2023,singh2024,sarkar2019,ding2019}.

The magnetic entropy change, $\Delta S_{\text{m}}(T)=\int_{0}^{T} \frac{C_{\text{m}}(T')}{T'},\mathrm{d}T'$, is shown in Fig.~\ref{yb_cp}(c). At zero magnetic field, the measurements extend down to 0.4~K,which is close to the peak in $C_{\text{m}}(T)$; hence, the magnetic entropy below this temperature cannot be determined experimentally from the present data. Nevertheless, the zero-field data recover approximately 46\% of $R\ln2$ by $\sim$20--30 K. We expect that the remaining entropy would be lost below 0.4~K. Similarly, for the 10 and 30~kOe measurements, the data were collected only down to 2~K, preventing a reliable estimation of the full entropy release, and hence these curves are omitted from the figure. For $H$ = 20~kOe, the maximum of $C_{\text{m}}(T)$ is around 2~K, while the data extend down to 0.4~K. In this case, the entropy change approaches the expected $R\ln2 = 5.763$~J~mol$^{-1}$~K$^{-1}$ within experimental uncertainty.

\begin{figure}[h]
	\centering
	\includegraphics[width=0.9\linewidth]{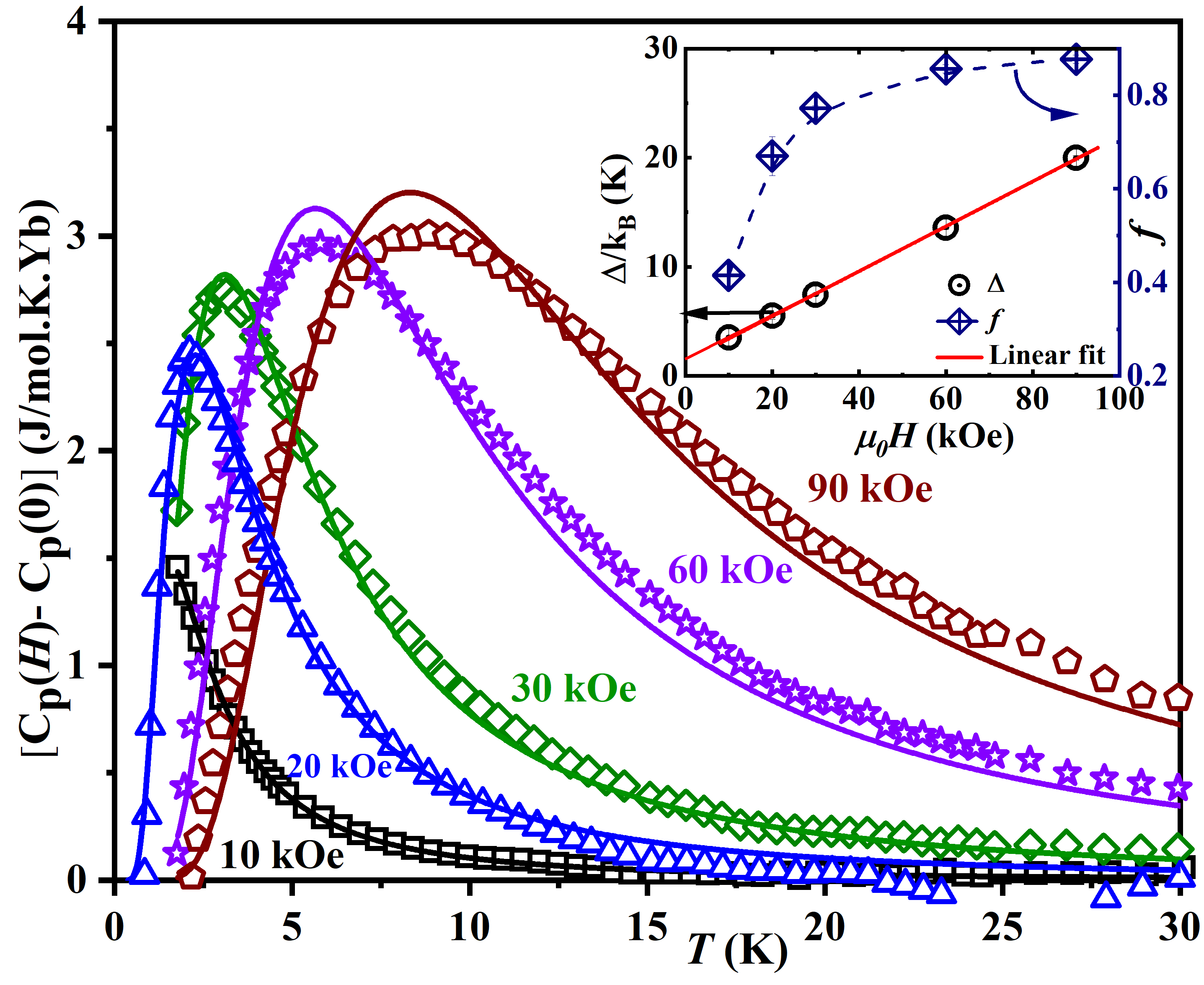}
	\caption{Schottky contribution to the specific heat of LCYBO at selected fields with two-level model fits (solid lines). Inset: field dependence of the energy gap $\Delta/k_{B}$ (left axis) and the fitted free-spin fraction $f$ (right axis).}
	\label{yb_csch}
\end{figure}

To quantify the field dependence, the magnetic contribution was modeled with a two-level Schottky expression appropriate for a $J_{\mathrm{eff}}=1/2$ doublet:

\begin{equation}
	C_{\mathrm{Sch}}(T,H) = fR \left(\frac{\Delta}{T^{2}}\right)
	\frac{\exp(\Delta/T)}{\big[1+\exp(\Delta/T)\big]^{2}},
	\label{eq:sch}
\end{equation}

where $f$ is the fraction of free spins per formula unit, $R$ is the universal gas constant, and $\Delta/k_B$ is the Zeeman splitting of the ground-state doublet. Fig.~\ref{yb_csch} shows the Schottky fits obtained from $C_{\mathrm{Sch}}(T,H)=C_{p}(H,T)-C_{p}(0,T)$. The fitted free-spin fraction $f$ increases with field and approaches a saturation value of about 86\%, indicating that the applied field progressively polarizes the moments and reduces the fraction of spins participating in correlated low-energy fluctuations. The extracted energy scale $\Delta/k_B$ varies linearly with $H$, following $\Delta/k_B = g(\mu_B/k_B)H$. The nonzero zero-field intercept, $\Delta(0)/k_B \simeq 1.35$~K, is observed here; note that zero-field gaps have been reported in many QSL candidate materials and might arise from correlated  Yb$^{3+}$ moments and/or disorder induced internal fields~\cite{kundu2020}.
The associated Schottky energy scale, $T_{\rm max} \approx 0.42,\Delta/k_B \sim 0.6$~K, is comparable to the temperature range where the low-temperature magnetic anomaly develops. This suggests that the observed broad feature in $C_{\rm mag}(T)$ may contain a Schottky-like contribution in addition to cooperative spin correlations. From the value $\Delta/k_B = 19.97$,K at 9,T, we obtain an effective $g$-factor $g \approx 3.3$, in good agreement with magnetization analysis.

\subsection{Nuclear Magnetic Resonance}

Nuclear magnetic resonance (NMR) provides a sensitive local probe of static and dynamic magnetism through the hyperfine coupling between nuclear and electronic moments. In LCYBO, the $^{7}$Li nuclei are coupled to the surrounding Yb$^{3+}$ moments via transferred hyperfine interactions through Li--O--Yb exchange pathways, making $^{7}$Li NMR an effective probe of the evolution of local spin correlations and slow magnetic fluctuations.

\begin{figure}[h]
	\centering
	\includegraphics[width=0.8\linewidth]{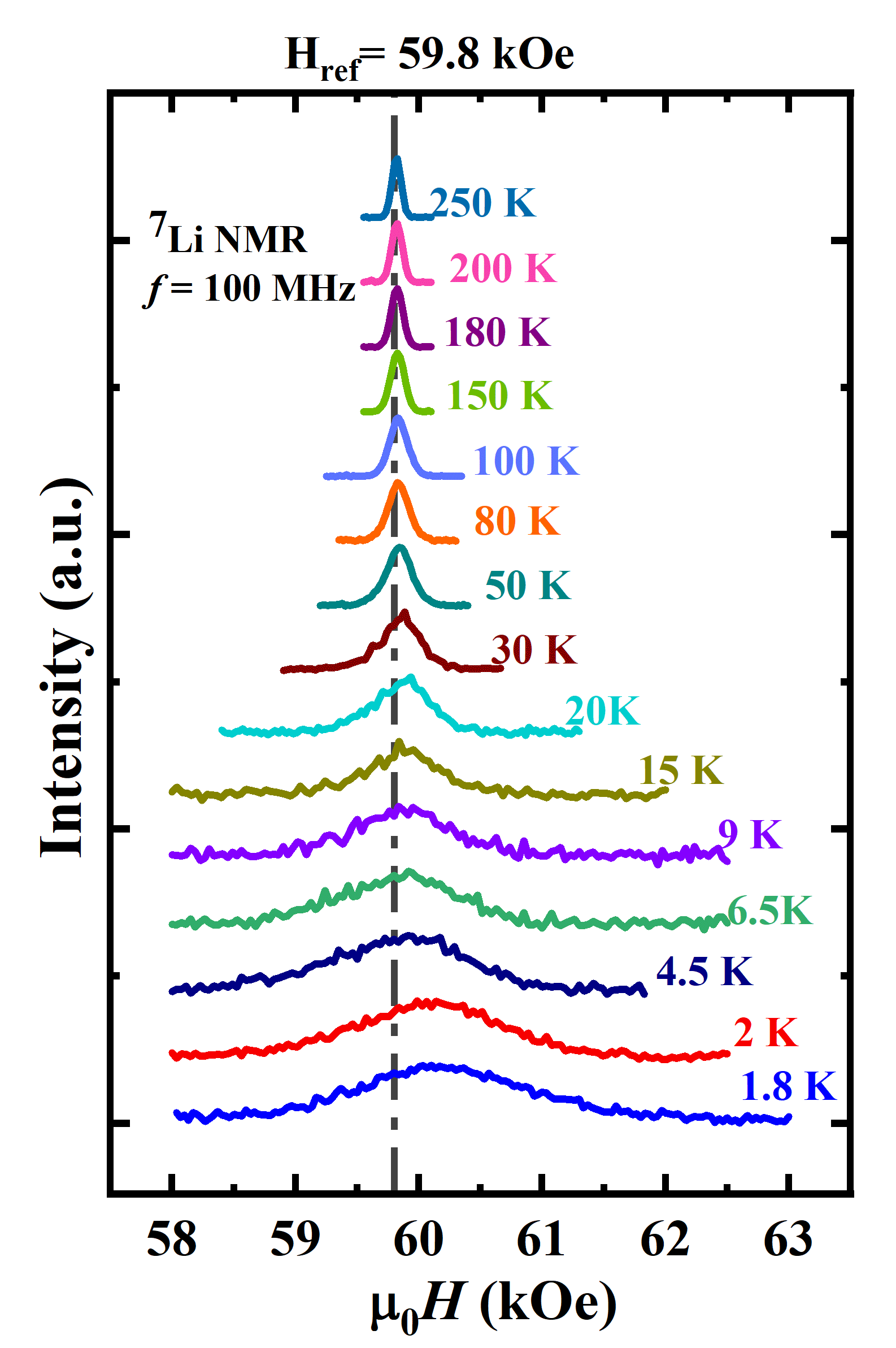}
	\caption{$^{7}$Li field-swept NMR spectra of LCYBO measured at a fixed frequency of 100\,MHz at selected temperatures. The dashed vertical line marks the reference field $H_{\rm ref}=59.8$\,kOe corresponding to zero shift. The progressive broadening and reduction in signal intensity upon cooling indicate increasing internal-field inhomogeneity and slowing spin dynamics associated with Yb$^{3+}$ moments.}
	\label{yb_nmr1}
\end{figure}

The $^{7}$Li nucleus has nuclear spin $I=3/2$, gyromagnetic ratio $\gamma/2\pi = 16.54607$\,MHz/T, and 100\% natural abundance. Since the resonance line broadens strongly on cooling, field-swept NMR spectra were measured at a fixed frequency of 100\,MHz over the temperature range 1.8--250\,K. The spectra were obtained using a standard spin-echo pulse sequence $(\pi/2-\tau-\pi)$. Representative spectra are shown in Fig.~\ref{yb_nmr1}. At high temperatures, the line is relatively narrow and nearly symmetric, whereas progressive broadening develops upon cooling, indicating an increasing distribution of internal hyperfine fields at the Li sites generated by correlated Yb$^{3+}$ moments. Below $\sim 20$\,K, the line becomes very broad and weak, consistent with slow spin dynamics and enhanced magnetic inhomogeneity.

Attempts to measure the spin--lattice relaxation time $T_1$ below $\sim 120$~K were unsuccessful because the substantial spectral broadening prevented uniform saturation of the entire NMR line. Under these conditions, only about $40\%$ of the total spectral intensity could be effectively saturated, resulting in incomplete recovery of the nuclear magnetization and unreliable fits.

\begin{figure}[h]
	\centering
	\includegraphics[width=1\linewidth]{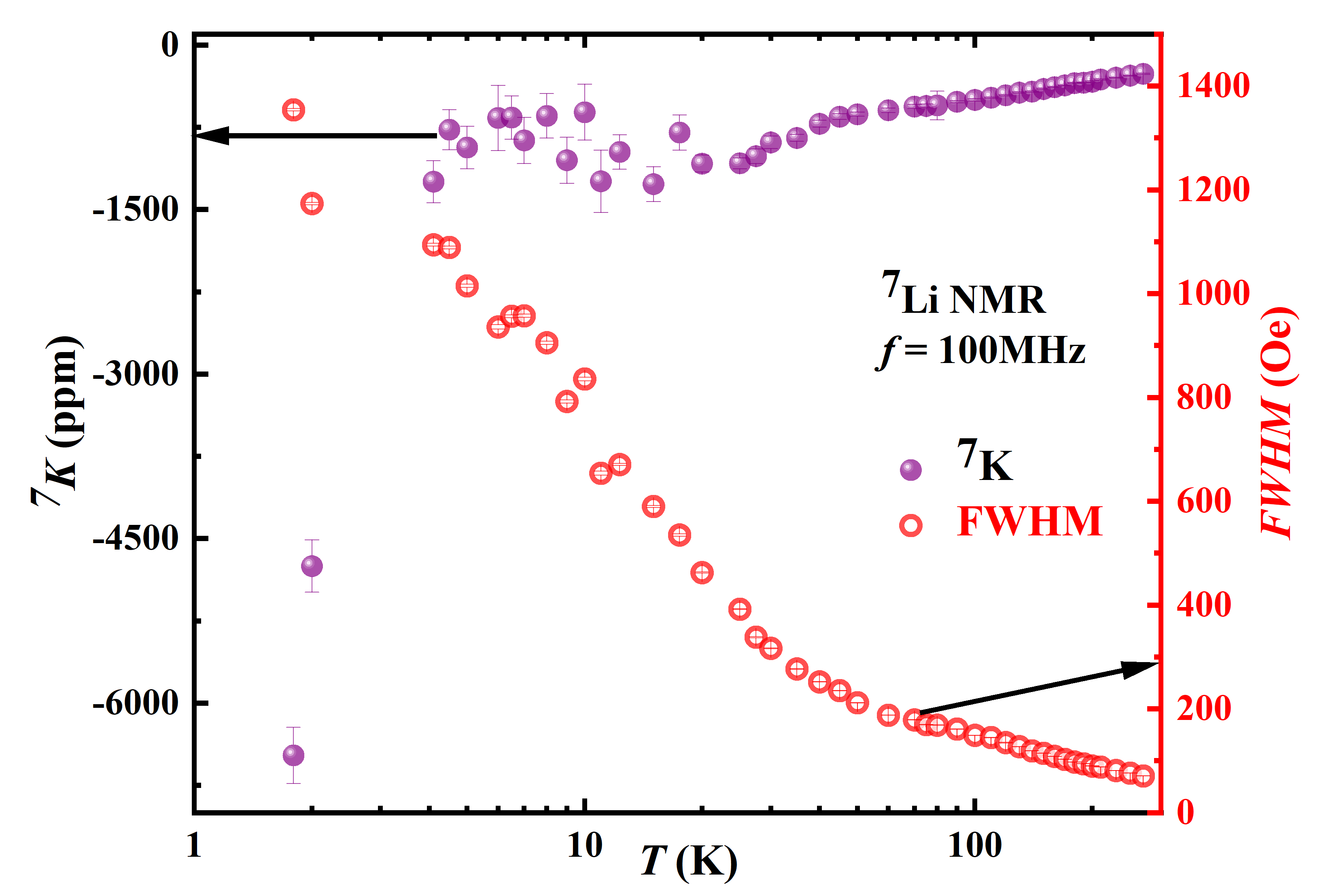}
	\caption{Temperature dependence of the $^{7}$Li Knight shift $K(T)$ (left axis) and spectral linewidth (FWHM, right axis) measured at 100\,MHz. The increasingly negative Knight shift and strong linewidth broadening below $\sim 50$\,K reflect the growth of the transferred hyperfine response and increasingly inhomogeneous low-energy spin dynamics.}
	\label{yb_nmr2}
\end{figure}

The Knight shift $K(T)$ [Fig.~\ref{yb_nmr2}, left axis] was extracted from the resonance field position using
\begin{equation}
	K(T) = \frac{H_{\mathrm{ref}} - H(T)}{H(T)},
	\label{eq:shift1}
\end{equation}

where $H(T)$ is the resonance field at temperature $T$, and $H_{\rm ref}=59.8$\,kOe is the reference field obtained from a nonmagnetic LiCl standard. The resonance field $H(T)$ was extracted by fitting the spectra to a Gaussian line shape at high temperatures, while the peak position was used in the broadened low-temperature regime. The temperature dependence of $K(T)$ is shown in Fig.~\ref{yb_nmr2} (left axis). The shift remains small and weakly temperature dependent above $\sim 50$\,K, indicating only modest transferred hyperfine fields in the paramagnetic regime. On cooling below $\sim 50$\,K, $K(T)$ becomes increasingly negative, followed by a pronounced downturn below $\sim 10$\,K. The negative shift indicates that the local hyperfine field at the Li site is opposite to the applied magnetic field, i.e., corresponding to a negative hyperfine coupling.

The full width at half maximum (FWHM), shown in Fig.~\ref{yb_nmr2} (right axis), exhibits complementary behavior. At high temperatures, the linewidth is narrow ($\sim 0.05$--0.1\,kOe) and only weakly temperature dependent. Below $\sim 50$\,K, the linewidth increases steadily, followed by a much stronger enhancement below $\sim 20$\,K, reaching $\sim 1.3$\,kOe at the lowest measured temperatures. Such substantial broadening reflects the development of quasi-static or slowly fluctuating internal fields and a broad distribution of local magnetic environments. The linewidth can also contain contributions from a distribution of transferred hyperfine couplings, anisotropic hyperfine interactions and powder averaging, Ca/Yb antisite disorder, local structural inhomogeneity, and dipolar fields. The broad linewidth is likely further enhanced by structural disorder arising from partial Yb/Ca antisite mixing and Li vacancies, which generate site-dependent hyperfine couplings.

To quantify the hyperfine coupling between the $^{7}$Li nuclei and Yb$^{3+}$ moments, we constructed a Clogston--Jaccarino plot of the Knight shift $K(T)$ versus the bulk susceptibility $\chi(T)$ measured at the same field. In the high-temperature paramagnetic regime, the data follow the linear relation
\begin{equation}
	K(T)=K_{0}+\frac{A_{\mathrm{hf}}}{N_{\mathrm A}\mu_{B}}\,\chi(T),
\end{equation}
where $K_{0}$ is the temperature-independent chemical/orbital shift and $A_{\mathrm{hf}}$ is the hyperfine coupling constant. The observed linearity confirms that the NMR shift is dominated by coupling to the Yb $4f$ moments. A linear fit restricted to this temperature range was used to determine $A_{\mathrm{hf}}$, where the intrinsic paramagnetic response dominates. At lower temperatures, deviations from linear $K$--$\chi$ scaling are observed, likely arising from extrinsic contributions to the bulk susceptibility, such as impurity-induced Curie tails and defect-related paramagnetic contributions, while the NMR shift continues to track the intrinsic local spin susceptibility.

\begin{figure}[ht]
	\centering
	\includegraphics[width=1\columnwidth]{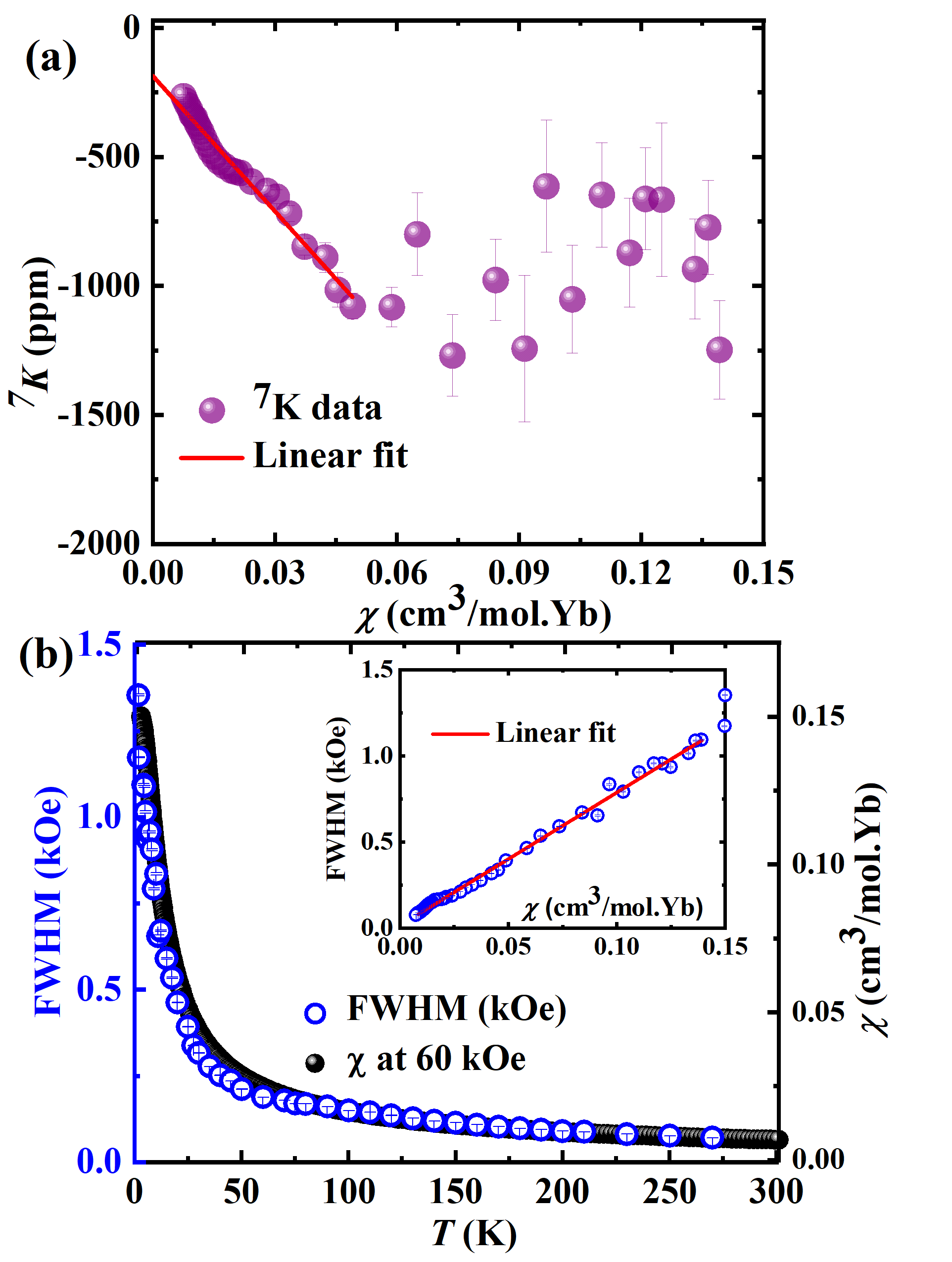}
	\caption{(a)~Clogston--Jaccarino plot of the $^{7}$Li Knight shift versus bulk susceptibility. The linear high-temperature scaling yields the hyperfine coupling $A_{\mathrm{hf}}$ and the temperature-independent shift $K_0$. (b)~Temperature dependence of the $^{7}$Li FWHM (left axis) and dc susceptibility measured at 60\,kOe (right axis). Inset: approximately linear scaling of FWHM with $\chi$, indicating a dominant susceptibility-dependent contribution to the linewidth.}
	\label{yb_nmr3}
\end{figure}

A linear fit to the high-temperature data [Fig.~\ref{yb_nmr3}(a)] yields $A_{\mathrm{hf}} = -98 \pm 3~\mathrm{Oe}/\mu_{B},$ $K_{0} = -186(13)~\mathrm{ppm}.$ The negative sign of 
$A_{\mathrm{hf}}$ indicates that the transferred hyperfine field at the Li site is antiparallel to the Yb moment.

Figure~\ref{yb_nmr3}(b) shows the temperature dependence of the full width at half maximum (FWHM) of the $^{7}$Li NMR line together with the dc susceptibility measured at 60~kOe. The linewidth closely tracks $\chi(T)$ over the full measured temperature range, indicating that susceptibility-dependent magnetic field distributions associated with the Yb$^{3+}$ moments make a major contribution to the observed linewidth. The inset of Fig.~\ref{yb_nmr3}(b) demonstrates an approximately linear relation between FWHM and $\chi$, while additional contributions from hyperfine-coupling distributions, anisotropic hyperfine interactions and powder averaging, structural disorder, and dipolar fields may also contribute to the absolute linewidth.

The temperature dependence of the $^{7}$Li NMR linewidth can be understood in terms of a temperature-independent dipolar contribution together with a susceptibility-dependent broadening arising from macroscopic field inhomogeneities in the powder sample. Using a Gaussian approximation, the full width at half maximum (FWHM) may be written as~\cite{mahajan1998}
\begin{equation}
	\Delta_{\mathrm{FWHM}}
	=
	2.35
	\sqrt{
		\langle \Delta \nu^{2}\rangle_{\mathrm{dip}}
		+
		\left(
		\frac{\gamma_n}{2\pi}B\chi H
		\right)^2
	},
	\label{eq:fwhm}
\end{equation}
where $\langle \Delta \nu^{2}\rangle_{\mathrm{dip}}$ is the second moment of the intrinsic dipolar broadening, $\chi$ is the measured molar susceptibility, $\gamma_n/2\pi = 16.546$~MHz/T is the gyromagnetic ratio of $^{7}$Li, $H$ is the applied magnetic field, and $B$ is the fractional root-mean-square deviation of the local magnetic field from the applied field arising from demagnetization and powder-packing effects. The numerical factor 2.35 converts the Gaussian standard deviation to the full width at half maximum.

The first term in Eq.~(\ref{eq:fwhm}) represents the second moment of the static dipolar interaction at the Li site. This quantity was calculated from the known crystal structure by performing the lattice sum over a large supercell containing 2197 translated unit cells, including both Yb--Li and Yb--Yb dipolar contributions~\cite{Abragam1961}. The corresponding lattice sums are $7.91\times10^{-3}\ \mathrm{\AA}^{-6}$ for the Yb--Li term and $2.38\times10^{-3}\ \mathrm{\AA}^{-6}$ for the Yb--Yb term, indicating that the Yb--Li contribution is the dominant microscopic source of dipolar broadening at the $^{7}$Li site. The total lattice sum is $1.03\times10^{-2}\ \mathrm{\AA}^{-6}$, which corresponds to an effective dipolar field scale $A_{\mathrm{dip}}=\left(2\pi g^{2}\mu_B^{2}\sum_j r_j^{-6}\right)^{1/2}\simeq 7.8$~kOe.

The second term in Eq.~(\ref{eq:fwhm}) is proportional to the bulk susceptibility and therefore accounts for the observed increase of linewidth on cooling. Experimentally, the FWHM versus $\chi$ data are well described by the linear relation $\Delta_{\mathrm{FWHM}}=m\chi+c$, with $m=7.91~\mathrm{kOe/(cm^{3}\,mol^{-1})}$ and $c=0.009$~kOe. Using $H=60$~kOe, we obtain an effective broadening parameter $B=m/H\simeq0.13$.

These results show that although the intrinsic dipolar field scale is sizeable, the observed linewidth is governed primarily by susceptibility-driven internal field distributions rather than purely static dipolar broadening. Such behavior is consistent with a frustrated Yb$^{3+}$ magnetic network, where geometrical frustration suppresses conventional long-range magnetic order and promotes short-range correlated spin textures. Upon cooling, the increase of $\chi(T)$ enhances the distribution of transferred hyperfine and dipolar local fields at the Li site, leading to the progressive broadening of the $^{7}$Li NMR line.

\paragraph*{\textbf{Spin--lattice relaxation.}}

\begin{figure}[ht]
	\centering
	\includegraphics[width=1\columnwidth]{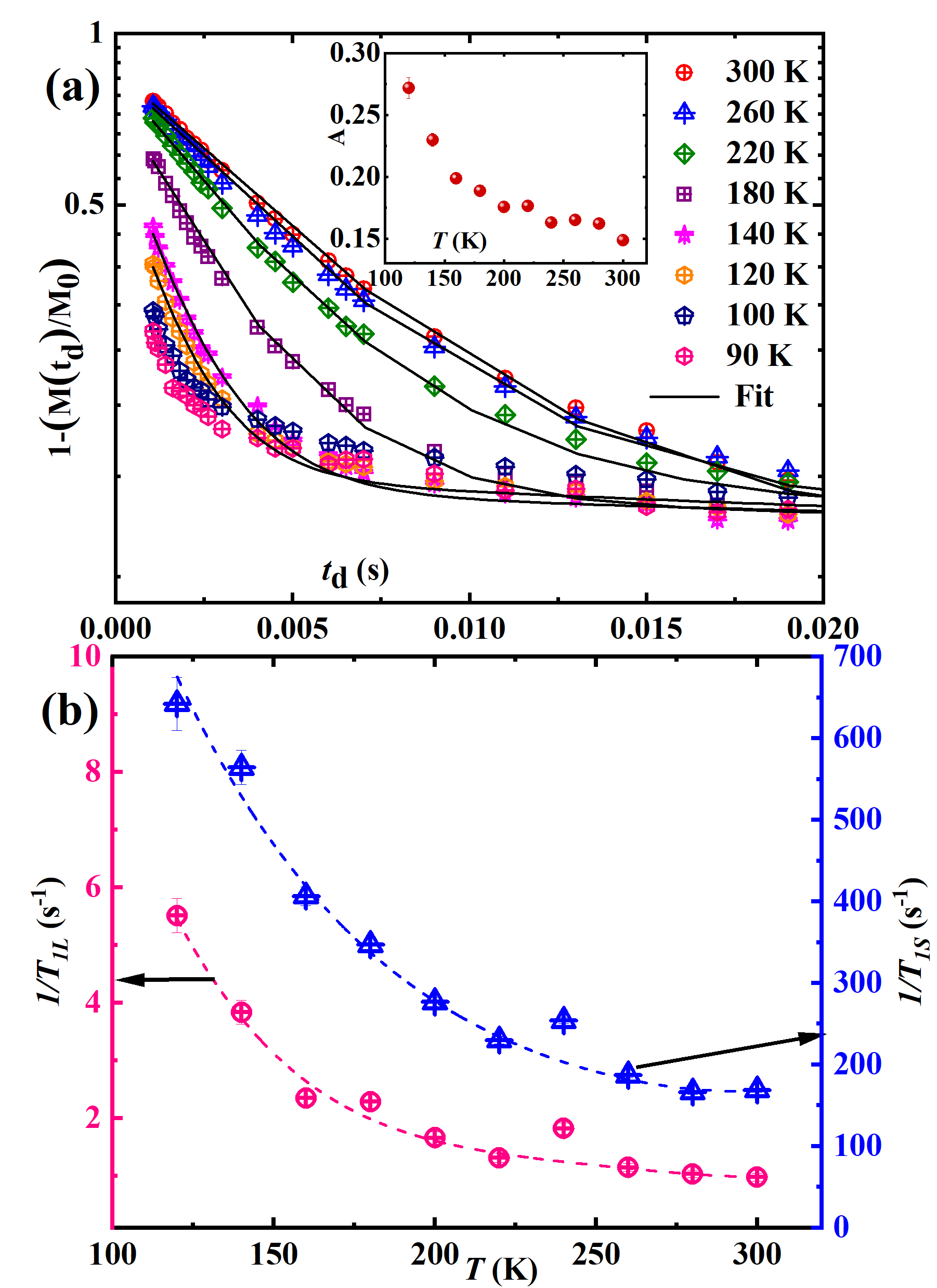}
	\caption{
		(a) Recovery of the longitudinal nuclear magnetization of $^{7}$Li at selected temperatures, plotted as $1 - M(t_d)/M_0$ versus delay time $t_d$. Solid lines represent fits using the double-exponential function given in Eq.~\ref{t1}, indicating the presence of two distinct spin--lattice relaxation channels. The inset shows the temperature dependence of the fractional weight $A$ associated with the slow relaxation component.  
		(b) Temperature dependence of the relaxation rates $1/T_{1L}$ (left axis) and $1/T_{1S}$ (right axis).The short component exhibits a significantly faster relaxation rate across the entire temperature range, consistent with enhanced low-frequency spin fluctuations in some disordered local environments, whereas the slow component may have a contribution from Li nuclei predominantly coupled to the fully occupied Yb(2) sublattice. The dashed lines are guides to the eye.}
	\label{yb_t1}
\end{figure}

The $^{7}$Li spin--lattice relaxation was measured as a function of temperature. Over the full accessible temperature range, the recovery of the longitudinal nuclear magnetization cannot be satisfactorily described by a single-exponential function. Instead, the data are well reproduced by a double-exponential form,

\begin{equation}
	M(t)=M_0\left[1-C\left(Ae^{-t/T_{1L}}+(1-A)e^{-t/T_{1S}}\right)\right],
	\label{t1}
\end{equation}

where $T_{1L}$ and $T_{1S}$ denote the long and short relaxation components, respectively, $A$ is the fractional weight of the slow component, and $C$ is a saturation factor determined by the saturation conditions. Representative recovery curves and fits are shown in Fig.~\ref{yb_t1}(a). The deviation from single-component recovery indicates a distribution of local magnetic environments and relaxation channels at the Li site.

The temperature dependences of $1/T_{1L}$ and $1/T_{1S}$ are shown in Fig.~\ref{yb_t1}(b). The faster rate remains substantially faster than the slower one throughout the measured range, indicating that a significant fraction of Li nuclei experience stronger fluctuating hyperfine fields at the NMR frequency scale. Both relaxation rates increase progressively on cooling from 300 to 120~K, evidencing a slowing down of Yb-spin fluctuations and the development of short-range magnetic correlations well above the low-temperature bulk anomaly. The relative weight $A$ of the slow component decreases with increasing temperature, from $\sim0.27$ near 120~K to $\sim0.15$ at 300~K, indicating a gradual redistribution between the two relaxation channels. Below approximately 120~K, pronounced spectral broadening prevents uniform saturation of the full NMR line, and only $\sim 40\%$ saturation of the longitudinal magnetization could be achieved near 100~K. In this regime, reliable determination of $T_1$ becomes increasingly difficult because a fraction of nuclei falls outside the effective detection window owing to very short transverse relaxation times $T_2$ and rapidly relaxing components.

The two-component relaxation is likely associated with disorder-induced variations in the local magnetic environment rather than multiple fully resolved crystallographic Li sites. Structural refinement reveals partial Ca/Yb antisite mixing at the Yb(1) and Yb(3) sites, with approximately $25\%$ Ca substitution at each site, while the Yb(2) site remains fully occupied by Yb$^{3+}$. Assuming random occupancy, four local environments are possible for Li nuclei, each characterized by different hyperfine couplings and exchange connectivities: (i) Yb(2)+Yb(1)+Yb(3) (56.25\%), (ii) Yb(2)+Yb(1) (18.75\%), (iii) Yb(2)+Yb(3) (18.75\%), and (iv) Yb(2) only (6.25\%). Environments (i)--(iii), which together account for approximately $94\%$ of all configurations, involve one or more partially occupied Yb1/Yb3 sites and may provide a plausible microscopic origin for the fast-relaxing component, owing to their different local magnetic environments and exchange connectivities. In contrast, the minority environment (iv), in which Li is coupled only to the fully occupied Yb2 sublattice, may provide a plausible microscopic origin for the long-relaxing component. Configurations involving partially occupied Yb(1)/Yb(3) sites may generate a broader distribution of fluctuating local fields and enhanced relaxation rates, whereas Li nuclei predominantly coupled to the Yb(2) network may exhibit comparatively slower relaxation.

Thus, the distributed spin--lattice relaxation is consistent with spatially inhomogeneous spin dynamics arising from structural disorder superimposed on a frustrated Yb magnetic network. The progressive loss of signal intensity and the inability to reliably determine $T_1$ at lower temperatures further support the development of slow and spatially heterogeneous magnetic fluctuations well above the low-temperature anomalies.
%-----------------------------------------------------------------

\begin{table*}[t]
	\centering
	\captionsetup{width=0.95\linewidth}
	\caption{Comparison of representative Yb$^{3+}$-based frustrated triangular-lattice magnets. Listed are the low-temperature Curie--Weiss temperature $\theta_{\mathrm{CW}}^{\mathrm{LT}}$, magnetic ordering temperature ($T_N$) or characteristic correlation scale ($T^{*}$), the ratio of interlayer to intralayer Yb spacing ($d_{\mathrm{inter}}/d_{\mathrm{intra}}$), the first excited crystal-electric-field (CEF) gap $\Delta_{\mathrm{CEF}}$, and the proposed magnetic ground state.}
	\label{tab:yb_comparison}
	\resizebox{\textwidth}{!}{
		\begin{tabular}{lccccccl}
			\hline
			\textbf{Compound} &
			\boldmath{$\theta_{\mathrm{CW}}^{\mathrm{LT}}$ (K)} &
			\textbf{$T_N$ / $T^{*}$ (K)} &
			\boldmath{$d_{\mathrm{inter}}/d_{\mathrm{intra}}$} &
			\boldmath{$\Delta_{\mathrm{CEF}}$ (K)} &
			\textbf{Ground state / Remarks} &
			\textbf{Ref.} \\
			\hline
			
			YbMgGaO$_4$
			& $-4$
			& 2.4
			& 2.47
			& 457
			& Triangular-lattice QSL candidate
			& \cite{li2015,li2020} \\
			
			NaYbO$_2$
			& $-10.3$
			& 1.0
			& 1.64
			& 320
			& Frustrated triangular antiferromagnet / QSL candidate
			& \cite{ding2019,bordelon2019} \\
			
			YbZnGaO$_4$
			& $-2.7$
			& 1.8
			& $\sim$2.4
			& 255
			& Disorder-induced frozen or glassy state
			& \cite{zhang2018,ma2020} \\
			
			NaBaYb(BO$_3$)$_2$
			& $-0.07$
			& $T_N \simeq 0.4$
			& 1.09
			& --
			& Weakly ordered triangular antiferromagnet
			& \cite{guo2019} \\
			
			YbBO$_3$
			& $-0.8$
			& $T_N \simeq 0.4$
			& 1.16
			& 647
			& Triangular-lattice antiferromagnet
			& \cite{Sala2023,somesh2023} \\
			
			\textbf{LiCaYb$_5$(BO$_3$)$_6$}
			& \textbf{-0.54}
			& \textbf{0.43}
			& \textbf{1.04}
			& \textbf{200}
			& \textbf{Weakly interacting distorted triangular-lattice magnet with disorder}
			& \textbf{(This work)} \\
			
			\hline
	\end{tabular}}
\end{table*}      

\section{Discussion}

The combined magnetic susceptibility, specific heat, magnetization, and $^{7}$Li NMR results establish LCYBO as a Yb$^{3+}$-based triangular-lattice magnet hosting a well-isolated effective $J_{\mathrm{eff}} = 1/2$ ground-state doublet. A weak anomaly observed near $T^{*} \simeq 0.43$~K suggests the development of a correlated low-temperature magnetic state. The low-temperature Curie--Weiss temperature, $\theta_{\mathrm{CW}}^{\mathrm{LT}} \approx -0.54$~K, indicates predominantly antiferromagnetic interactions between the Yb$^{3+}$ moments. Analysis of the magnetic susceptibility using a two-level crystal-electric-field (CEF) model yields a first excited CEF level near 200~K, confirming that the low-temperature magnetism is governed by a well-separated Kramers doublet.

The specific heat data further support this picture. At high temperatures, the response is dominated by lattice contributions, while a magnetic contribution develops below $\sim 5$\,K, consistent with the development of short-range spin correlations. The weak anomaly near $T^{*}$ gradually evolves into a broad Schottky-like feature under applied magnetic fields, suggesting that the zero-field anomaly may contain contributions from low-energy magnetic correlations in addition to a Schottky-like component. Fits to the field-dependent Schottky contribution yield an effective $g$ factor of approximately 3.3, in good agreement with the magnetization analysis. Furthermore, the magnetic entropy recovered at 20~kOe approaches nearly $98\%$ of $R\ln2$, consistent with a well-isolated doublet ground state. In contrast, at zero field, the full magnetic entropy cannot be assessed within the present measurement range down to 0.4~K, as the contribution from temperatures below 0.4~K remains experimentally inaccessible.

The magnetic properties of LCYBO are closely connected to its crystal structure. The compound crystallizes in the hexagonal $P6_522$ space group, where three inequivalent Yb sites form slightly distorted triangular layers. Although the three Yb sites are crystallographically inequivalent, they form a common connected magnetic network rather than independent magnetic subsystems; their distinct local environments may result in different exchange couplings and $g$ tensors. The inequivalent Yb–Yb pathways may give rise to nonuniform and anisotropic exchange interactions. Together with the triangular geometry, these inequivalent exchange interactions can introduce magnetic frustration and compete with the development of conventional long-range magnetic order. In addition, partial Ca/Yb antisite disorder and fractional Li occupancy introduce local randomness in the magnetic exchange network, further modifying the exchange pathways and magnetic environments. Thus, frustration arising from the distorted triangular geometry, exchange anisotropy associated with the inequivalent Yb sites and bond geometry, and disorder-induced exchange randomness coexist in LCYBO. These combined effects may contribute to the broad correlated regime, broadened magnetic anomalies, and reduced entropy recovery observed in the system.

The $^{7}$Li NMR measurements provide microscopic insight into the local magnetic environment and spin dynamics. At high temperatures, the Knight shift scales linearly with the bulk susceptibility, yielding a hyperfine coupling constant $A_{\mathrm{hf}} \simeq -98$~Oe/$\mu_B$. Upon cooling below $\sim 20$~K, both the Knight shift and linewidth increase substantially, reflecting the progressive development of short-range magnetic correlations and quasistatic internal fields. The nearly linear scaling between the linewidth and $\chi(T)$ suggests that the broadening is dominated primarily by transferred hyperfine and dipolar fields arising from correlated Yb moments rather than purely structural disorder. The estimated dipolar contribution is also consistent with the magnitude of local fields expected from nearby rare-earth moments.

The spin--lattice relaxation measurements further indicate an inhomogeneous magnetic environment. The recovery of the nuclear magnetization could not be described by a single exponential function and instead required a double-exponential form, implying a distribution of local relaxation channels. This behavior is consistent with multiple Li environments arising from structural disorder, inequivalent hyperfine couplings, or spatially varying magnetic correlations. Below $\sim 120$~K, the substantial broadening of the NMR spectrum prevented homogeneous saturation of the full line, precluding reliable determination of $1/T_1$. The loss of measurable low-temperature relaxation data is consistent with substantial spectral broadening arising from increasingly distributed internal magnetic fields at the Li site.

Comparison with other Yb$^{3+}$-based frustrated magnets places LCYBO in an interesting intermediate regime between conventional ordered systems and highly frustrated quantum-disordered materials. Canonical triangular-lattice quantum spin-liquid candidates such as YbMgGaO$_4$~\cite{li2015,li2020}, NaYbO$_2$~\cite{ding2019,bordelon2019}, and YbZnGaO$_4$~\cite{zhang2018,ma2020} remain magnetically disordered down to the lowest measured temperatures despite sizable antiferromagnetic interactions. In contrast, LCYBO exhibits a weak low-temperature anomaly near $0.43$~K in thermodynamic measurements. However, the absence of a sharp $\lambda$-type transition, the broad correlated regime extending above $T^{*}$, and the persistent NMR linewidth broadening indicate that the low-temperature state may not be a conventional long-range-ordered phase.

Overall, LCYBO realizes a correlated effective $J_{\mathrm{eff}} = 1/2$ triangular-lattice antiferromagnet with weak signatures of low-temperature static magnetism. The interplay of triangular geometry, anisotropic exchange interactions, and disorder is consistent with the observed correlated low-energy magnetic state, with only weak signatures of static ordering at very low temperatures. Future microscopic probes such as $\mu$SR and neutron scattering will be essential for determining whether static long-range magnetic order is established, quantifying the static magnetic volume fraction, and clarifying the nature of the residual low-energy spin dynamics.

\section{Conclusion}

High-quality polycrystalline LiCaYb$_5$(BO$3$)$6$ (LCYBO) was successfully synthesized, and Rietveld refinement confirmed a single-phase hexagonal $P6_522$ structure. Magnetic susceptibility, magnetization, specific heat, and $^{7}$Li NMR measurements establish LCYBO as a frustrated Yb$^{3+}$ triangular-lattice magnet with a well-isolated effective $J{\mathrm{eff}} = 1/2$ ground-state doublet. The low-temperature Curie--Weiss temperature, $\theta{\mathrm{CW}}^{\mathrm{LT}} \approx -0.54$~K, indicates weak antiferromagnetic interactions, while crystal-electric-field analysis places the first excited level near 200~K. Specific heat data reveal the onset of short-range magnetic correlations below $\sim 5$~K and a weak anomaly near $T^{*} \simeq 0.43$~K. Under applied magnetic fields, this feature evolves into a broad Schottky-like anomaly associated with Zeeman splitting of the ground-state doublet, yielding an effective $g$ factor of about 3.3. The magnetic entropy recovered at 20~kOe approaches nearly $98\%$ of $R\ln2$, consistent with a well-isolated Kramers doublet ground state.

$^{7}$Li NMR measurements provide microscopic evidence for correlated magnetism and distributed local magnetic environments. The Knight shift scales linearly with bulk susceptibility, giving a hyperfine coupling constant of approximately $-98$~Oe/$\mu_B$, while strong linewidth broadening at low temperatures reflects transferred hyperfine and dipolar fields from correlated Yb moments. The spin--lattice relaxation shows a two-component recovery, suggesting a distribution of local relaxation channels likely arising from structural disorder and inequivalent hyperfine environments. Together, the results suggest that the combined effects of the geometric frustration, exchange anisotropy, and structural randomness may suppress conventional magnetic ordering and contribute to an extended correlated low-temperature regime. These findings position LCYBO as a promising Yb-based platform for exploring the crossover between weakly ordered and quantum-disordered triangular-lattice magnetism.

\section*{Acknowledgments}
We acknowledge the use of various Central Facilities at IIT Bombay. M.J. would like to acknowledge the funding support for Ph.D. fellowship DST-Inspire fellowship ((IF180933) from the DST, Govt. of India. S.N. would like to acknowledge the funding support for Chanakya Postdoctoral fellowship (CPDF/2021-22/01) from the National Mission on Interdisciplinary Cyber Physical Systems, of the DST, Govt. of India through the I-HUB Quantum Technology Foundation. This work was partially supported by the Deutsche Forschungsgemeinschaft (DFG) within the Transregional Collaborative Research Center TRR 360 ``Constrained Quantum Matter", Project No. 492547816 (Augsburg, Munich, Stuttgart, Leipzig). 
 
%%---------------------------------------------
%                     Reference Page
%%---------------------------------------------
\bibliography{ref}

\end{document}